\documentstyle[preprint,aps,floats,epsfig]{revtex}
\tighten

\newcommand{\be}{\begin{equation}}
\newcommand{\ee}{\end{equation}}
\newcommand{\bea}{\begin{eqnarray}}
\newcommand{\eea}{\end{eqnarray}}
\newcommand{\bml}{\begin{mathletters}}
\newcommand{\eml}{\end{mathletters}}

\begin{document}
\preprint{hep-th/0003109}

\title{Brane worlds: the gravity of escaping matter}
\author{Ruth Gregory$^{1}$, 
Valery A. Rubakov$^{2}$ and Sergei M. Sibiryakov$^{2}$  \\
{\small \em ~$^1$ Centre for Particle Theory, Durham University,
South Road, Durham, DH1 3LE, U.K.}\\
{\small \em {~}$^2$ Institute for Nuclear Research of the Russian Academy of Sciences,}\\
{\small \em 60th October Anniversary prospect, 7a, Moscow 117312, Russia.}
}
\maketitle
\begin{abstract}
In the framework of a five-dimensional model with one 3-brane and
an infinite extra dimension, we discuss a process in which matter
escapes from the brane and propagates into the bulk to arbitrarily 
large distances.  An example is a decay of a particle of mass $2m$ 
residing on the brane into two particles of mass $m$ that leave 
the brane and accelerate away. We calculate, in the linearized theory,
the metric induced by these particles on the brane. This metric does
not obey the four-dimensional Einstein equations and corresponds to a
spherical gravity wave propagating along the four-dimensional future
light cone. The four-dimensional space-time left behind the spherical
wave is flat, so the gravitational field induced in the brane world by
matter escaping from the brane disappears in a causal way.
\end{abstract}



\newpage
\section{Introduction and summary}
It has been found recently \cite{Randall:1999vf}
that four-dimensional gravity may emerge as a low energy effective
theory in models with non-compact extra dimensions (see
also Refs \cite{Cohen:1999ia,Gregory:1999gv}). One ingredient of
this scenario is a relatively old  
\cite{Rubakov:1983bb,akama,visser}, and recently revived
\cite{Arkani-Hamed:1998rs,Antoniadis:1998ig,Randall:1999ee} 
idea that ordinary matter may reside
on a 3-brane embedded in higher-dimensional space. Another key
point is the existence in a class of models 
of a bound state of a multi-dimensional
graviton which is localized near the brane \cite{Randall:1999vf}.
Even though gravity at short distances is higher-dimensional,
this bound state dominates gravitational interactions
in the brane world at large distances and gives rise to
four-dimensional behavior of the gravity force experienced by matter
residing on the brane. 
Modifications of this scenario include models with a
metastable graviton bound state \cite{Charmousis:1999rg,Gregory:2000jc} or
more than one very light four-dimensional graviton \cite{Kogan:1999wc};
these models predict violations of the four-dimensional Newton's law
at ultra-large distances as well.

If the extra dimensions are non-compact, it is conceivable that energy
may leak from the brane into the bulk. As an example, field theoretic
models of 3-branes can be viewed as defects in higher dimensions 
(domain walls in $(4+1)$-, cosmic strings in $(4+2)$-dimensions etc.). 
In these models, highly energetic ordinary particles are able to 
leave the brane and propagate into the bulk. Another possibility is that 
besides ordinary matter, there exist particle species that are not trapped 
to the brane at all; pair creation of these particles would also lead to the
transfer of energy from the brane to the bulk. The same role may be
played by Kaluza--Klein gravitons whose interactions with the brane
matter are weak but non-vanishing.

At first glance, the possibility that energy is carried away from the 
brane to arbitrarily large distances in the bulk seems, from the
four-dimensional point of view, to be in conflict with locality, 
causality, and the
four-dimensional Newton's law (``Gauss' law of general relativity''): 
changing mass in a finite region of three-dimensional space would seem 
to result in an instantaneous changes of the gravitational potential 
everywhere in this space. By this argument, an observer living on the 
brane would immediately realize that energy had been emitted from the 
brane to the bulk, no matter how far away this event occurred.

Off hand, one may think of several possible
resolutions of this apparent paradox. Three of them 
are as follows:

(i) One may recall that in $(1+1)$ dimensions, energy non-conservation
is in fact consistent with locality, causality and the long-range
character of forces analogous to gravity 
\cite{Rubakov:1997tw,Rubakov:1997wb}. Even though the reasons for this
property are peculiar to $(1+1)$ dimensions, one may wonder whether
a similar consistency may be possible in $(3+1)$ dimensions.

(ii) The bulk geometry in the Randall--Sundrum (RS)  model 
\cite{Randall:1999vf} and its analogs
is anti-de Sitter, so that particles leaving the brane get accelerated
away from the brane.  Their energy increases as they move away, 
and one may wonder whether this effect could compensate for
the decrease of the strength of gravitational interactions of these
particles with matter on the brane. In that case, the
mass  measured through gravitational forces by an observer residing on
the brane would remain constant, and the particles continuously
accelerating in the bulk would behave, from the four-dimensional point of
view, as dark matter particles of fixed mass which participate in
gravitational interactions in a standard way.

(iii) Finally,  gravity in the RS model is guaranteed to be effectively
four-dimensional only as far as interactions of matter residing on the
brane are concerned. No argument implies that the effective
four-dimensional description is valid for gravity induced by bulk
matter. In other words, the gravitational field induced on the brane
by particles moving in the bulk need not obey the four-dimensional
Einstein equations, so the above discussion of the conflict between
causality and Newton's law  may not apply.

In this paper we decide between these possibilities by calculating, in
the linear theory about the RS background, the gravitational field of a
pair of particles escaping from the brane and propagating in the bulk.
We show that the most exotic option (i) has nothing to do with the RS
model.  The possibility (ii) would be realized if the graviton bound state
were the only relevant mode, i.e., if the zero mode approximation were
reliable. This may be viewed as a consistency check: in the zero-mode
approximation gravity is effectively four-dimensional irrespective
of the position of its source in extra dimensions, so the
four-dimensional gravitating mass must be conserved. 

However, the zero-mode approximation is in fact not adequate to the problem in
question, and the Kaluza--Klein (KK) tower of gravitons plays an important
role. When KK states are included in the analysis, the outcome is
option (iii). The final picture is that the particles moving away from
the brane produce a spherical gravity wave on the brane, which expands
in three-dimensional space with the  speed of light (or almost the speed of
light). The four-dimensional space-time left behind this spherical wave is flat,
whereas in front of this wave the four-dimensional metric is of the
usual Schwarzschild asymptotic form, in accord with causality. The
spherical gravitational wave itself does not obey the four-dimensional
Einstein equations, so the gravitational effects 
of particles moving in the bulk are intrinsically higher-dimensional,
even if these effects are measured by an observer residing on the
brane.

\section{ The physical set up}

The Randall--Sundrum model contains a 3-brane
with tension $\sigma > 0$  embedded in 
five-dimensional space-time. The bulk cosmological 
constant between the branes, $\Lambda$, is
negative and tuned in such a way that the intrinsic geometry on the brane is
that of flat four-dimensional Minkowski spacetime. The solution to the
five-dimensional Einstein equations is
\be
ds^2=a^2(z)\eta_{\mu\nu}dx^{\mu}dx^{\nu}-dz^2
\ee
where
\be
a(z) =  e^{-k|z|} 
\ee
Here $\mu,\nu = 0,1,2,3$ and $\eta_{\mu\nu}$ is the Minkowski metric.
The constant $k$ is related to $\sigma$ and $\Lambda$ as follows:
$\sigma=\frac{3k}{4\pi G_5}$, $\Lambda=-\sigma k$, where $G_5$ is
the five-dimensional Newton constant. The four-dimensional hypersurfaces
$z=const.$ are flat, the five-dimensional space-time to the right of the
brane is a part of anti-de Sitter (adS) space.

It is sometimes convenient to introduce another coordinate,
\be
\zeta = \frac{1}{k} e^{kz}
\label{defzeta}
\ee
in terms of which the background metric is conformally flat,
\be
ds^2 = \frac{1}{k^2\zeta^2}\left( \eta_{\mu\nu} dx^{\mu}dx^{\nu} 
- d\zeta^2 \right)
\ee

Consider now a process  in which some matter is emitted from the
brane, and then freely moves in the bulk adS background.
Let us assume for simplicity that this matter is symmetric
with respect to $z \to -z$;
this will enable us to take symmetric metric perturbations 
and effectively consider only the space to the right of the brane.
An example which we discuss throughout this paper
is two  particles of mass $m$
which are emitted from the brane at time $t=0$ 
along the line ${\bf x} = 0$
in opposite directions in $z$ with zero initial velocities.
The coordinates of these particles obey the geodesic equations 
in adS. It is straightforward to see that the world line of
a particle right to the brane is described as follows,
\be
{\bf x}_c(t) = 0\,,\;\;\;\;
z_c(t) = \frac{1}{2k} \ln (1 + k^2 t^2)
\label{part}
\ee
In terms of the coordinate $\zeta$, this means
\be
\zeta_c(t) = \sqrt{k^{-2} + t^2}
\label{zetasource}
\ee
The particle accelerates towards $z\to \infty$, and at
$t \gg k^{-1}$ its world line approaches the light cone,
$\zeta = t$. Similar observations have been made independently
in Ref.\cite{Volovich}.

The energy-momentum tensor of this particle,
\be
\hat{T}^{ab} = \frac{m}{\sqrt{-g}} \frac{dx^a}{ds} 
\frac{dx^b}{dt} \delta ({\bf x}-{\bf x}_c(t)) \delta (z -z_c(t))
\label{tt2}
\ee
has the following non-vanishing components,
\bml\label{tti}\bea
T_{zz} &=& \frac{m}{a^3} \frac{v^2}{\sqrt{1-v^2}} 
\delta (z-z_c(t)) \delta ({\bf x})
\label{tt1} \\
T_{z0} &=& - \frac{m}{a^2} \frac{v}{\sqrt{1-v^2}} 
\delta (z-z_c(t)) \delta ({\bf x}) \\
T_{00} &=&  \frac{m}{a} \frac{1}{\sqrt{1-v^2}} 
\delta (z-z_c(t)) \delta ({\bf x})
\label{tt3}
\eea\eml
where
\be
v = \frac{\dot{z_c}}{a(z_c)} = \frac{kt}{\sqrt{1+k^2t^2}}
\ee
Hereafter we consider tensors with all lower indices as basic ones;
quantities with and without hats have upper indices raised by full adS
and Minkowski metrics respectively. As an example,
$T^{\mu}_{\nu} = \eta^{\mu\lambda} T_{\lambda \nu}$,
$\hat{T}^{\mu}_{\nu} = a^{-2}\eta^{\mu\lambda} T_{\lambda \nu}$.

Equations (\ref{tt2}) and (\ref{tti}) are valid at $t>0$; before that
the energy-momentum tensor is concentrated on the brane.
We will assume for definiteness that at $t <0$, the source on the
brane is a point-like particle with mass $2m$. The physical picture is
that this brane particle decays at $t=0$ into two particles of equal
mass, one of which moves according to eq.\ (\ref{part}) and another which is
its mirror image.


The linearized five-dimensional Einstein equations in RS background
have been considered in a number of papers, see, e.g., Refs.\ \cite
{Randall:1999vf,Charmousis:1999rg,Lykken:1999nb,Garriga:1999yh,Giddings:2000mu}.
In particular, the advantages and disadvantages of the Gaussian Normal
(GN) gauge have been emphasized 
\cite{Garriga:1999yh,Giddings:2000mu,Charmousis:1999rg}. 
As we are interested in the gravitational effects {\it on the brane}
of a particle moving {\it outside the brane}, we find it convenient to work 
entirely in GN coordinates, however, this will mean that there will not
be a global coordinate system which is GN once a matter source has been
introduced.  In a GN frame one has
\be
g_{zz} = -1\,,\,\,\, g_{z\mu} = 0   
\label{5*}
\ee
and  the linearized theory is described by the metric
\be
ds^2 = a^2(z) \eta_{\mu\nu} dx^{\mu} dx^{\nu} 
+ h_{\mu\nu}(x,z)dx^{\mu}dx^{\nu} - dz^2  
\ee
The linearized Einstein equations have the following form,
\bml\bea
\delta R_{zz} &=& 8\pi G_5\theta_{zz} \label{1}\\
\delta R_{z\mu} &=& 8\pi G_5 \theta_{z\mu} \label{2}\\
\delta R_{\mu\nu} - 4k^2 h_{\mu\nu} &=&  8\pi G_5\theta_{\mu\nu}
\label{3}
\eea\eml
where
\bml\bea
\theta_{zz} &=& \left( \frac{2}{3} T_{zz} + \frac{1}{3a^2} 
T^{\lambda}_{\lambda} \right) \label{TeZZ}\\
\theta_{z\mu} &=&  T_{z\mu}\\
\theta_{\mu\nu} &=& \left( T_{\mu\nu}
- \frac{1}{3} \eta_{\mu\nu}T^{\lambda}_{\lambda} + 
\frac{a^2}{3} \eta_{\mu\nu} T_{zz} \right) \label{TeMN}
\eea\eml
and
\bml\bea
\delta R_{zz} &=& - \left(\frac{h'}{2a^2}\right)'  \\
\delta R_{z\mu} &=& \left[\frac{1}{2a^2}(h^{\nu}_{\mu,\nu}
-h_{,\mu} )\right]' \\
\delta R_{\mu\nu}&=&\frac{1}{2}h_{\mu\nu}'' + 2k^2 h_{\mu\nu}
-\left(k^2 h + \frac{k}{2} h'\right)\eta_{\mu\nu} \nonumber \\
&&  +\frac{1}{2a^2} (2 h^{\lambda}_{(\mu,\nu)\lambda} 
- h^{\;\;\;\;,\lambda}_{\mu\nu,\lambda} - h_{,\mu\nu} )
\eea\eml 
Hereafter $h = h^{\mu}_{\mu} \equiv \eta^{\mu\nu} h_{\mu\nu}$.

We will explicitly consider times $t > 0$ when the source is in the bulk.
Since there is no matter on the brane, the metric should obey
the Israel junction condition on the brane, which, due to the $Z_2$ symmetry 
reads
\be
h_{\mu\nu}' + 2k h_{\mu\nu} = 0      
\label{junc1}
\ee
Let us first solve eq.\ (\ref{1}). Its general solution is
\be
h(z,x) = -\frac{ 8\pi G_5}{k} \left[ a^2(z) 
\int_{z}^{\infty}~dz'~\theta_{zz}(z') 
-  \int_{z}^{\infty}~dz'~a^2(z')\theta_{zz}(z') \right]
+ a^2(z) C(x) + D(x)
\label{deter}
\ee
From the junction condition, $h' + 2k h = 0$, we find
\be
D(x) = -\frac{ 8\pi G_5}{k}  \int_{0}^{\infty}~dz'~a^2(z')\theta_{zz}(z')
\label{D}
\ee
Consider now eq.\ (\ref{3}). Let us introduce
\be
\xi_{\mu} = h_{\mu,\lambda}^{\lambda} - \frac{1}{2} h_{,\mu}
\label{defx}
\ee
This function can be found from eq.\ (\ref{2}), but its explicit form
will be irrelevant. With the definition (\ref{defx}), 
eq.\ (\ref{3}) becomes
\be
\frac{1}{2}h_{\mu\nu}'' - 2k^2 h_{\mu\nu}
- \frac{1}{2a^2} \Box h_{\mu\nu} =
8\pi G_5\theta_{\mu\nu} + \left(k^2 h + \frac{k}{2} h'\right) \eta_{\mu\nu}
- \frac{1}{2a^2}(\xi_{\mu,\nu} + \xi_{\nu,\mu})
\label{hfull}
\ee
Hence, one can write
\be
h_{\mu\nu} = \bar{h}_{\mu\nu} + (u_{\mu,\nu} + u_{\nu,\mu})
\ee
where $u_{\mu}$ accounts for $\xi_{\mu}$ in eq.\ (\ref{hfull}),
and $\bar{h}_{\mu\nu}$ obeys the following equation,
\be
\frac{1}{2}\bar{h}_{\mu\nu}'' - 2k^2 \bar{h}_{\mu\nu}
- \frac{1}{2a^2} \Box \bar{h}_{\mu\nu} =
8\pi G_5\theta^{eff}_{\mu\nu}
\label{htrunc}
\ee
with
\be
8\pi G_5\theta^{eff}_{\mu\nu} =  8\pi G_5\theta_{\mu\nu}
+ \left(k^2 h + \frac{k}{2} h'\right) \eta_{\mu\nu}
\label{Teeff}
\ee
The last term in this equation is known explicitly from eqs.\  (\ref{deter})
and (\ref{D}):
\be
\left(k^2 h + \frac{k}{2} h'\right) 
= - 8\pi G_5k \int_0^z~dz'~a^2(z') \theta_{zz}(z')
\label{addterm}
\ee
We are interested in the induced metric on the brane,
\be
h_{\mu\nu} (z=0) = \bar{h}_{\mu\nu}(z=0) + 
(u_{\mu,\nu} + u_{\nu,\mu})(z=0)
\ee
The last term in this equation is a pure gauge in the four-dimensional
brane world and has no effect on the motion of matter confined to the brane,
so we may omit it. In effect, we have to solve eq.\ (\ref{htrunc})
and then find $\bar{h}_{\mu\nu}$ on the brane, i.e., at $z=0$.

It is worth looking at the additional term (\ref{addterm}) in the case of
the point-like particle moving in the bulk. We have
\be
\left(k^2 h + \frac{k}{2} h'\right) =
- 8\pi G_5k \theta(z-z_c(t)) a^2(z_c(t)) \Phi_{zz}(x)
\label{nonlocal}
\ee
where $\Phi_{ab}$ is defined by
\be
\theta_{ab} (z,x) = \delta (z-z_c(t)) \Phi_{ab} (x)
\label{Teeff2}
\ee
Explicitly
\bml\bea
\Phi_{zz}(x) &=& \frac{m}{a^3(z_c) \sqrt{1-v^2}}\left(\frac{2}{3}v^2
+ \frac{1}{3}\right) \delta ({\bf x})
\label{zz}\\
\Phi_{00}(x) &=& \frac{m}{a(z_c) \sqrt{1-v^2}}\left(\frac{1}{3}v^2
+ \frac{2}{3}\right) \delta ({\bf x})
\label{00}\\
\Phi_{ij}(x) &=& \frac{m}{a(z_c) \sqrt{1-v^2}}\left(-\frac{1}{3}v^2
+ \frac{1}{3}\right) \delta ({\bf x}) \delta_{ij}
\label{ij}
\eea\eml
A special feature of the additional term (\ref{nonlocal}) is that 
this is a non-local expression with a ``string'' extending from $z=z_c$
to $z=\infty$.  This string is of course a gauge artifact due to the
breakdown of the brane GN gauge as we pass the particle in the
$z$-direction, and represents a caustic of the normal geodesic congruence
used to define the affine parameter $z$. In fact, 
a similar string is also characteristic to 
the asymptotic four-dimensional Schwarzchild solution transformed into a gauge
which is GN with respect to an arbitrarily chosen 2-plane.
This artifact is easily removable (as noted in \cite{Giddings:2000mu})
via a construction reminiscent of 
the Wu-Yang gauge patching for a Dirac monopole \cite{WY}. One 
introduces an additional GN gauge patch to the right of the accelerating 
particle, valid for $z>z_c(t)-\epsilon\tan^{-1}
|{\bf x}|$, with the brane GN patch being valid for $z<z_c(t)+\epsilon\tan^{-1}
|{\bf x}|$. The gauge transformation on the overlap is readily seen to be
a bulk analog of the Garriga-Tanaka transformation:
\be
\epsilon_\mu = - {\epsilon_z(x)\over2k} \;\; ; \hskip 1cm
\partial^2 \epsilon_z = 8\pi G_5 a^2\left(z_c(t)\right) \Phi_{zz}(x)
\ee
One could always choose a harmonic bulk gauge ($\nabla^ah_{ab} 
- {1\over2}h_{,b}=0$), however, the computation of the Green's function
in this gauge is cumbersome and not particularly illuminating, therefore
we simply use a GN gauge. Indeed,
since we are primarily interested in the metric induced on the brane,
we stick with the brane GN system, the string singularity outside 
the brane not leading to any inconvenience.

\section{Solution of the linearized problem}

\subsection{First trial:  zero mode approximation}

We are going to treat the parameter $k$ of the RS model as microscopic,
and are interested in the induced metric on the brane, 
$\bar{h}_{\mu\nu}(x, z=0)$, at  distance scales large compared to
$k^{-1}$. In particular, we consider the region of the
four-dimensional space-time such that $r\equiv |{\bf x}| \gg k^{-1}$, 
$t \gg k^{-1}$. The solution to eq.\  (\ref{htrunc}) involves
the retarded Green's function of the operator 
\be
\frac{1}{2}\frac{d^2}{dz^2} - 2k^2 - \frac{1}{2a^2} \Box
\label{opera}
\ee
with boundary conditions enforcing the junction property
(\ref{junc1}). This Green's function is expressed in terms of the
eigenfunctions of the corresponding one-dimensional problem
\cite{Randall:1999vf,Garriga:1999yh},
\be
G_R(x-x',z,z')=
4ka^2(z)a^2(z')D_0(x-x')+
2\int_0^\infty dmu_m(z)u_m(z')D_m(x-x')
\label{generalGf}
\ee
where $D_0$ and $D_m$ are retarded Green's functions of massless and
massive scalar fields in four flat dimensions, and
\be
u_m(z)=\sqrt{\frac{m}{k}}
\frac{J_1(m/k)Y_2(m\zeta)-Y_1(m/k)J_2(m\zeta)}
{\sqrt{J_1(m/k)^2+Y_1(m/k)^2}}
\label{mode}
\ee
Here $\zeta(z)$ is defined by eq.\  (\ref{defzeta}), $J_n$ and $Y_n$ are
the Bessel functions.
The first term in eq.\ (\ref{generalGf}) is the contribution of the
bound state of the five-dimensional graviton (the zero mode), whereas
the second term comes from the continuous KK spectrum.
Note that our expression for the Green's function,
eq.\ (\ref{generalGf}), differs from that of
Refs.\ \cite{Randall:1999vf,Garriga:1999yh} by an overall factor of 4.
One factor of 2 has to do with our form of the operator (\ref{opera}), and
the other is due to symmetry $z \to -z$ and effectively accounts for 
the {\it two} particles moving left and right from the brane.

If the source in eq.\ (\ref{htrunc}) were on the brane, the
long-distance behaviour of the induced metric would be governed by
the zero mode contribution,
\be
G_R^{zm}(x-x',z,z')=
4ka^2(z)a^2(z')D_0(x-x')
\label{0modec}
\ee
As our first trial, let us boldly use the zero mode approximation
(\ref{0modec}) in the problem at hand. As mentioned in the Introduction,
this approximation is {\it not} adequate in our case, but rather
provides a non-trivial consistency check.

In the zero mode approximation, 
the metric induced on the brane is effectively determined 
by the following equation,
\be
    - \Box \bar{h}_{\mu\nu} (x) =  8\pi G_N\tau_{\mu\nu} (x)
\label{boxbox}
\ee
where $G_N=kG_5$ is the four-dimensional Newton constant and
\be
    \tau_{\mu\nu}(x) = 4 \int_0^{\infty}~dz~a^2(z)
     \theta^{eff}_{\mu\nu}(z)
\ee
In the case of a point particle, the right hand side of this equation is
straightforward  to
evaluate,
\bml\bea
\tau_{00} &=& 2m \frac{a(z_c(t))}{\sqrt{1-v^2}}
\delta(\bf{x}) \label{t00}\\
\tau_{ij} &=& 2 m\frac{a(z_c(t))}{\sqrt{1-v^2}}
\delta({\bf x}) \delta_{ij}
\label{tij}
\eea\eml
Making use of eq.\ (\ref{part}) one finds
\be
\frac{a(z_c(t))}{\sqrt{1-v^2}} = 1
\ee
Therefore, eq.\ (\ref{boxbox}) coincides with the linearized Einstein
equation in four dimensions with a static source of mass $2m$. As
discussed in the Introduction, this is consistent with the general
property that gravity in the zero mode approximation is always
effectively four-dimensional, irrespective of the position of its
source in the fifth dimension.

\subsection{Full treatment}

To obtain the correct expression for the induced metric on the brane,
we have to consider the complete Green's function
(\ref{generalGf}). Is is convenient to work in the coordinate
representation, and make use of the explicit form of the retarded
scalar
Green's function in four flat dimensions \cite{BSh},
\be
	D_m(x)=- \frac{1}{2\pi}\theta(t)\delta(\lambda^2) +
	\frac{m}{4\pi\lambda}\theta(t-|{\bf x}|)J_1(m\lambda),
	~~~\lambda=\sqrt{t^2-|{\bf x}|^2}
\label{coordGf}
\ee
At $m=0$ (i.e., in the zero mode case), only the first term in this
expression survives.

The first term in eq.\ (\ref{coordGf}) is independent of $m$. Now, 
one recalls that the functions $u_m(z)$, with the zero mode
included, form a complete set, so that
\be 
2ka^2(z)a^2(z') + \int u_m(z)u_m(z')dm = a^2(z)\delta(z-z')
\ee
We see that the contribution of the zero mode to the five-dimensional
Green's function (\ref{generalGf}) is cancelled out at $z\neq z'$
by KK states. Since we are interested in the metric {\it on} the 
brane induced by the particle {\it outside} the brane, we 
set $z\neq z'$ and obtain
\be
G_R(x;z,z')=
\int_0^\infty ~dm~ \frac{m}{2\pi\lambda}\theta(t-|{\bf x}|)
J_1(m\lambda)u_m(z)u_m(z')\, ,\;\;\;\; z\neq z'
\label{bef}
\ee
To analyze the long-distance physics, we consider the regime
$t,|{\bf x}|,\zeta' \gg k^{-1}$ (recall that $\zeta \sim t$
at the position of the source, see eq.\ (\ref{zetasource})).
Then the main contribution to  the Green's function
comes from modes with small $m$, that is
$m\ll k$. Thus, we retain the terms in eq.\ (\ref{mode}) which are 
leading order in $m/k$ (not assuming that $m\zeta'$ is small)
and obtain for the Green's function with one argument on the brane and
another off the brane
\be
G_R (x;0,z')= - \frac{\theta(t-|{\bf x}|)}{2\pi k\lambda} 
\int_0^\infty
dm~m^2 J_1(m\lambda) J_2(m\zeta')
\label{aft1}
\ee
The integration here is performed by making use of the following
relations, 
\bea
\int_0^\infty dx~x^2 J_1(\alpha x)J_2(\beta x)&=&
\left(\frac{\partial^2}{\partial \beta^2}-\frac{1}{\beta}
\frac{\partial}{\partial \beta}\right)
\int_0^\infty dx J_1(\alpha x)J_0(\beta x) \nonumber \\
&=&
\left(\frac{\partial^2}{\partial \beta^2}-\frac{1}{\beta}
\frac{\partial}{\partial \beta}\right)
\frac{1}{\alpha}\theta(\alpha-\beta)
\eea
In this way we find
\be
G_R (x;0,z')= - \frac{\theta(t-|{\bf x}|)}{2\pi k \lambda^2}
\left(\delta'(\lambda-\zeta')+\frac{1}{\zeta'}\delta(\lambda-\zeta')
\right)
\label{aft}
\ee
The Green's function $G_R(x-x';0,z')$ is concentrated on the 
five-dimensional light cone, $\lambda = \zeta'$, i.e.,
\be 
(t - t')^2 - ({\bf x} - {\bf x}')^2 - \zeta^{\prime~2} = 0
\ee
This feature may (or may not) be 
an artifact of our coarse-graining: if we were able to
resolve short distances, the Green's function might spread over
the region of size $k^{-1}$ inside the light cone.

To obtain the metric induced on the brane,
we have to integrate the Green's function (\ref{aft})
with $\theta_{\mu\nu}^{eff}$,
\bea
\bar{h}_{\mu\nu}(x)
\equiv \bar{h}_{\mu\nu}(x, z=0)
&=& 8\pi G_5\int d^4x' \int_0^\infty\! dz' G_R(x-x';0,z')
\theta_{\mu\nu}^{eff}(x',z')  \nonumber \\
&=& 8\pi G_5\int d^4x' \int_{1/k}^\infty\! \frac{d\zeta'}{k\zeta'} 
G_R(x-x';0,\zeta') \theta_{\mu\nu}^{eff}(x',\zeta')
\label{met}
\eea
The source here is given by  eqs.\ (\ref{Teeff}), (\ref{nonlocal}), 
(\ref{Teeff2}),
\be
\theta_{\mu\nu}^{eff}(x,\zeta)=k\zeta\delta(\zeta-\zeta_c(t))
\Phi_{\mu\nu}(x)-\frac{1}{k\zeta_c^2(t)} \eta_{\mu\nu}
\theta(\zeta-\zeta_c(t))\Phi_{zz}(x)
\label{thet}
\ee
Even though this source contains the non-local ``string'', the
integrand in eq.\ (\ref{met}) is in effect local. This is due to the 
particular structure of the Green's function (\ref{aft}),
\be
G_R (x;0,z') =
\frac{\theta(t-|{\bf x}|)}{2\pi k \lambda^2}
\zeta'\frac{\partial}{\partial \zeta'}
\left(\frac{1}{\zeta'}\delta(\lambda-\zeta')
\right)
\ee
Upon integrating by parts in eq.\ (\ref{met}), the second term in
eq.\ (\ref{thet}) becomes a delta-function of $(\zeta - \zeta_c(t))$,
i.e., only the region $\zeta \approx \zeta_c(t)$ in fact contributes
to this integral. This of course had to  be the case, as the ``string'' is a
gauge artifact.

In the regime considered, $t \gg k^{-1}$,  one has $\zeta_c(t) =t$,
and the leading terms in $\Phi_{ab}$ are
(see eqs.\  (\ref{zz},\ref{00},\ref{ij}))
\bml\bea
\Phi_{zz}&=&m(kt)^4\delta({\bf x})\\
\Phi_{00}&=&m(kt)^2\delta({\bf x})\\
\Phi_{ij}&=&{m\over3}\delta_{ij}\delta({\bf x})
\label{Ph}
\eea\eml
Substituting these expressions  into eq.\  (\ref{thet}) we get
\bml\bea
\theta_{00}^{eff}(x,\zeta) &=& m[(kt)^3\delta(\zeta-t)-k^3t^2
\theta(\zeta-t)]\delta({\bf x})\\
\theta_{ij}^{eff}(x,\zeta) &=& mk^3t^2
\theta(\zeta-t)\delta_{ij}\delta({\bf x})
\eea\eml
The integration in eq.\  (\ref{met}) is now straightforward. We end up with
\bea
\bar{h}_{00}(x)&=&-4G_Nm~\frac{2t^2 - r^2}{t^3}\, , \nonumber\\
\bar{h}_{ij}(x)&=&-4G_Nm~\frac{1}{t}\delta_{ij}\, ,
\;\;\;\;\;\; \;\;\;\;\; t-r > 0
\label{61}
\eea
Here we stress the fact that these expressions are valid inside
the four-dimensional future light cone, $t-r > 0$, as is clear from
the explicit $\theta$-function in eq.\  (\ref{bef}). Outside this light cone,
the induced metric is determined by the source existing at
$t<0$. With our model for this source (particle of mass $2m$ on the
brane at rest at ${\bf x} = 0$), the induced metric outside the light
cone is the asymtotic Schwarzschild solution (see,
e.g. Ref.\ \cite{Garriga:1999yh}  for explicit derivation in RS model),
which in an appropriate gauge reads
\bea
\bar{h}_{00}(x)&=&-4G_Nm~\frac{1}{r}\, , \nonumber \\
\bar{h}_{ij}(x)&=&-4G_Nm~\frac{1}{r}\delta_{ij}\, , \;\;\;\;\; t-r < 0
\label{62}
\eea
We see that the induced metric (\ref{61}), (\ref{62}) is continuous
on the four-dimensional light cone, but its derivatives are not.
Because of the latter property, the induced metric does not obey the
(linearized) four-dimensional Einstein equations. It describes a
spherical gravity wave propagating in four dimensions with the speed
of light. Again, within our approximation we do not resolve distances
of order $k^{-1}$, so the wave may actually spread over the region of
this size.

The four-dimensional space-time on the brane left behind the spherical
wave is in fact flat. Indeed, a  coordinate transformation on the brane
\be
\delta t =-4G_Nm\left(\frac{r^2}{4t^2} + \ln{t}\right),~~~
\delta x^i = -4G_Nm~\frac{x^i}{2t}
\ee
reduces the four-dimensional metric perturbation (\ref{61}) to
$\bar{h}_{\mu\nu} = 0$ inside the light cone $t-r >0$. The
gravitational field induced on the brane by matter escaping into the
bulk finally disappears, this process occuring in a causal way.

\section{ Discussion}

We have shown that within linearized perturbation theory, the metric
on the brane does indeed react to the `loss' of the sources in
the bulk in an intrinsically five-dimensional fashion, a spherical
shock wave expanding outwards from the moment of emission leaving
behind flat space. It is tempting to ask if something similar
could perhaps be obtained by an appropriate use of the far field
Schwarzschild-adS solution
\be
ds^2 = \left ( 1 + k^2 \rho^2 - {\mu\over k^2\rho^2} \right ) d\tau ^2-
\left ( 1 + k^2 \rho^2 - {\mu\over k^2\rho^2} \right )^{-1} d\rho^2
- \rho^2 d\Omega_{I\!I\!I}^2
\label{schads}
\ee
In the absence of the mass term, $\mu$, the transformation between
the brane coordinates and the spherical coordinates is
\bml\bea
\rho &=& {1\over 2k\zeta} \left [ (\zeta^2-k^{-2})^2 + k^2({\bf x}^2 - t^2)^2 + 
2k^2\zeta^2
({\bf x}^2 - t^2) + 2(t^2 + {\bf x}^2) \right ] ^{1/2} \\
\tan k\tau &=& {2t\over k} \left [ k^{-2}
+ \zeta^2 + {\bf x}^2 - t^2 \right ]^{-1} \\
\tan \chi &=& {2 |{\bf x}|\over k} \left [ \zeta^2
+ {\bf x}^2 - t^2 - k^{-2} \right ] ^{-1}
\eea\eml
where the parametrisation of S$^3$ in the spherical system is the direct
generalisation of the S$^2$ one, rather than Euler angles, and
the angular $\theta,\phi$ variables coincide with the angular variables
on the brane-world. Notice the periodic relation of the $\tau$-variable
to the brane coordinates, this signifies that the spherical coordinates
represent in fact the universal covering space of adS, whereas the brane
exists in a single patch. In a sense, there are an infinite family
of branes existing in the maximally extended spacetime, as discussed
in Chamblin, Hawking and Reall \cite{CHR}. These branes start off `planar',
become parabolic, then return to planar again, oscillating indefinitely
in the maximally extended spherical coordinates. From the perspective of
the brane spacetime, $\rho=0$ corresponds to the geodesic trajectory
(\ref{part}). It is tempting therefore to ask whether we
can, by modifying our spacetime to Schwarzschild-adS, come up with a
metric corresponding to the particle accelerating away from the brane,
at least at large spherical $\rho$-coordinate. From the perspective
of the metric (\ref{schads}), this would correspond to placing a brane
at some angular coordinate $\chi = \chi(t,\rho)$.

In Ref. \cite{CHR}, it was shown that it was not possible to
find a static trajectory, $\chi(\rho)$, which might correspond to
a particle sitting on the brane, however, it was suspected that this
was due to the non-accelerating nature of the black hole -- a more
appropriate exact metric, such as some sort of C-metric would 
be a better candidate. These expectations were partly backed up by
the lower-dimensional calculation of \cite{EHM}. In our case however, 
the accelerating particle in the bulk translates into a non-accelerating
$\rho=0$ geodesic in the spherical spacetime (\ref{schads}) so we might hope
that a time dependent brane trajectory will work. 
Unfortunately, as we show in the appendix,  it
is not possible even to find a time-dependent trajectory corresponding
to a particle accelerating in the bulk. The reason why this approach
fails (as well as that in \cite{CHR}) becomes apparent once we
think of the difference in the causal structure of the adS and
Schwarzschild-adS spacetimes. In the former, we have an infinite
family of oscillating branes, whereas the latter has simply one copy of
the brane. The trajectory of the brane therefore cannot be a simple perturbation
of the pure adS trajectory, which oscillates periodically in $\tau$.
Indeed, if one were to try this approach, one would be trying to consider the
cumulative effect of the attractive central potential generated by
the Schwarzschild source over an infinite number of oscillations!
It becomes clear that (\ref{schads}) is an inappropriate
metric to use in this context when one tries to use it as a far-field
approximation, i.e.\ assuming that the matter source is extended in such
a fashion as to avoid an event horizon. A brane-world observer sitting 
outside the domain of influence of this accelerating particle 
(the past light cone of $t=-k^{-1}$ for example) should see no effect,
however, the metric (\ref{schads}) still induces a nonzero perturbation in this
region, which is of course due to the infinitely extended nature of 
the maximal adS spacetime. This behaviour should be contrasted with
the causal behaviour of the metric perturbation derived above.

\vskip 1cm
\begin{center}
{\bf Acknowledgments}
\end{center}

We would like to thank Victor Berezin, Christos Charmousis,
Ed Copeland, Savas Dimopoulos,
Sergei Dubovsky, Gia Dvali, Dmitry Gorbunov,
Maxim Libanov, John March-Russell, Lisa Randall, Simon Ross
and Sergei Troitsky
for useful discussions. R.G.\ and V.R.\ acknowledge the
hospitality of the Isaac Newton Institute for Mathematical Sciences,
where this work was begun.
R.G.\ is supported by the Royal Society, and V.R. and S.S by
the Russian Foundation for Basic Research, grant 990218410.

\section*{Appendix}

In this appendix, we show that it is not possible to use the Schwarzschild-adS
solution (\ref{schads}) to construct a spacetime corresponding to a brane
with an accelerating particle in the bulk.
To do this, we need only two of the four
seperate Israel conditions appropriate to the hypersurface defined by
the brane:
\be
X^a = (\tau, \rho, \chi(\tau,\rho), \theta, \phi)
\ee
with normal
\be
n_a = ({\dot\chi},\chi',-1,0,0)/n
\ee
where $n^2 = = 1/\rho^2  + A^2 \chi^{\prime 2}  - {\dot\chi}^2A^{-2}$, 
writing $A^2=g_{tt}$ for convenience. The $\theta$ and $\phi$
Israel conditions for the brane are identical, and give
\be
{( \cos\chi - \rho A^2\chi'\sin\chi ) \over n \rho^2 \sin\chi} = k
\label{angis}
\ee
In order to get the remaining Israel conditions, rather than working
with the fundamental forms of the brane hypersurface, it is easier
to generalise the technique used in Ipser and Sikivie, \cite{IS},
and use the normal jump in the parallel derivatives of the unit vectors
corresponding to the remaining time and space-like directions on the 
brane:
\be
u^a = {\dot X}^a / |{\dot X}|\; , \;\;\;\;
v^a = X^{a\prime} / |X'|
\ee
which satisfy
\be
n_a \nabla_u u^a = -k \; , \;\;\; 
n_a \nabla_v v^a = k \; , \;\;\; 
n_a \nabla_v u^a = n_a \nabla_u v^a = -k u^av_a
\ee
This last relation giving
\be
-{\dot\chi}' + {\dot\chi} \left ( {A'\over A} - A^2 \rho \chi^{\prime2}
-{1\over \rho} \right ) = k \rho^2 {\dot\chi} \chi' n
\label{crosis}
\ee
Combining (\ref{angis}) and (\ref{crosis}) requires
\be
{\dot\chi}\sin\chi  = {A\over \rho} f(k\tau) \Rightarrow
\cos\chi = -{A\over \rho} F(k\tau) + C(\rho)
\ee
where $F= \int d\tau f$.  But then (\ref{angis}) implies
\be
f^2 + F^2 - 2{AF\over k\rho} \left( C + \rho C'\right ) \left (
1 + {2\mu\over k^4\rho^4} \right )+ A^2 \left( C + \rho C'\right )^2
+ {\mu\over k^4\rho^4} \left ( 3F^2 - C^2 - \rho^2A^2C^{\prime2} \right ) = 1
\ee
It is not difficult to see that since $f,F$ are functions of $\tau$
this cannot be satisfied once $\mu\neq0$. For $\mu=0$ the solution
is of course the adS brane trajectory: $f=-\sin k\tau$, $C=-1/k\rho$.

We should also note that this approach is distinct from that of 
Kraus and Ida \cite{KI}, who use the Schwarzschild-adS spacetime
to obtain a {\it cosmological} brane world, i.e.\ a homogeneous
universe with an evolving scale factor, which we would not expect 
to obtain from a localised particle accelerating in the bulk.


\begin{thebibliography}{99}
\bibitem{Randall:1999vf}
L.~Randall and R.~Sundrum,
Phys.\ Rev.\ Lett.\  {\bf 83} (1999) 4690
[hep-th/9906064].

\bibitem{Cohen:1999ia}
A.~G.~Cohen and D.~B.~Kaplan,
Phys.\ Lett.\  {\bf B470} (1999) 52
[hep-th/9910132].


\bibitem{Gregory:1999gv}
R.~Gregory,
``Nonsingular global string compactifications,''
hep-th/9911015.


\bibitem{Rubakov:1983bb}
V.~A.~Rubakov and M.~E.~Shaposhnikov,
Phys.\ Lett.\  {\bf B125} (1983) 136.

\bibitem{akama}
K.~Akama,
in {\it Gauge Theory and Gravitation. 
Proceedings of the International Symposium,
Nara, Japan, 1982}, eds. K.~Kikkawa, N.~Nakanishi and
H.~Nariai (Springer--Verlag, 1983).

\bibitem{visser}
M.~Visser,
Phys.\ Lett.\  {\bf B159} (1985) 22.

\bibitem{Arkani-Hamed:1998rs}
N.~Arkani-Hamed, S.~Dimopoulos and G.~Dvali,
Phys.\ Lett.\  {\bf B429} (1998) 263;
Phys.\ Rev.\  {\bf D59} (1999) 086004

\bibitem{Antoniadis:1998ig}
I.~Antoniadis, N.~Arkani-Hamed, S.~Dimopoulos and G.~Dvali,
Phys.\ Lett.\  {\bf B436} (1998) 257

\bibitem{Randall:1999ee}
L.~Randall and R.~Sundrum,
Phys.\ Rev.\ Lett.\  {\bf 83} (1999) 3370


\bibitem{Charmousis:1999rg}
C.~Charmousis, R.~Gregory and V.~A.~Rubakov,
``Wave function of the radion in a brane world,''
hep-th/9912160.

\bibitem{Gregory:2000jc}
R.~Gregory, V.~A.~Rubakov and S.~M.~Sibiryakov,
``Opening up extra dimensions at ultra-large scales,''
hep-th/0002072.

\bibitem{Kogan:1999wc}
I.~I.~Kogan, S.~Mouslopoulos, A.~Papazoglou, G.~G.~Ross and J.~Santiago,
``A three three-brane universe: New phenomenology for the new millennium?,''
hep-ph/9912552.

\bibitem{Rubakov:1997tw}
V.~A.~Rubakov,
Phys.\ Rev.\  {\bf D56} (1997) 3523

\bibitem{Rubakov:1997wb}
V.~A.~Rubakov,
in {\it Proceedings of International Seminar on Quantum Field Theory
and High Energy Physics, Protvino, Russia, 1997},
gr-qc/9711055.

\bibitem{Volovich} W.~Mueck, K.~S.~Viswanathan and I.~V.~Volovich,
``Geodesics and Newton's Law in Brane Backgrounds,''
hep-th/0002132.

\bibitem{Lykken:1999nb}
J.~Lykken and L.~Randall,
``The shape of gravity,''
hep-th/9908076.

\bibitem{Garriga:1999yh}
J.~Garriga and T.~Tanaka,
``Gravity in the brane-world,''
hep-th/9911055.

\bibitem{Giddings:2000mu}
S.~B.~Giddings, E.~Katz and L.~Randall,
``Linearized Gravity in Brane Backgrounds,''
hep-th/0002091.

\bibitem{WY}
T.~T.~Wu and C.~N.~Yang,  
Phys.\ Rev.\ {\bf D12} (1975) 3845. Nucl.\ Phys.\ {\bf B107} (1976) 365.


\bibitem{BSh} N.~N.~Bogoliubov and D.~V.~Shirkov,
 {\it Introduction to Theory of Quantized Fields}, 
(Moscow, Nauka, 1976).

\bibitem{CHR}  A.\ Chamblin, S.\ W.\ Hawking and H.\ S.\ Reall, 
Phys.\ Rev.\ {\bf D61} (2000) 065007. [hep-th/9909205].

\bibitem{EHM} R.\ Emparan, G.\ T.\ Horowitz and R.\ C.\ Myers,
JHEP 0001 (2000) 007. [hep-th/9911043].

\bibitem{IS} J.\ Ipser and P.\ Sikivie, 
Phys.\ Rev.\ {\bf D30} (1984) 712.

\bibitem{KI} P.\ Kraus, JHEP 9912 (1999) 011. [hep-th/9910149].\\
D.\ Ida, ``Brane world cosmology'', gr-qc/9912002.

\end{thebibliography}
\end{document}